\documentclass[pre,reprint,twocolumn,amsmath,10pt,amssymb,aps,superscriptaddress,showkeys,showpacs]{revtex4-1}\input epsf
\usepackage{bm}

\usepackage[toc,page]{appendix}
\usepackage{soul}
\usepackage{graphicx}
\usepackage{makeidx}
\usepackage{chemarr}
\usepackage{multirow}
 \usepackage{color}
 \usepackage[normalem]{ulem}
\definecolor{darkgreen}{rgb}{0,0.6,0}
 \definecolor{orange}{rgb}{0.99,0.257,0}
\newcommand{\be}{\begin{equation}}
\newcommand{\ee}{\end{equation}}
\newcommand{\ba}{\begin{eqnarray}}
\newcommand{\ea}{\end{eqnarray}}

\def\ie{{\it i.e}~}

\def\D{\Delta}

\def\la{\langle}
\def\ri{\rangle}

\def\epsilon{\varepsilon}

\def\ie{i.e.}

\def\beqr{\begin{eqnarray}}
\def\eqnr{\end{eqnarray}}
\def\beq{\begin{equation}}
\def\bc{\begin{center}}
\def\ec{\end{center}}
\def\eqn{\end{equation}}

\usepackage{amsmath}

\usepackage{graphicx}
\usepackage{dcolumn}
\usepackage{bm}

\begin{document}

\preprint{AIP/123-QED}

\title{Influence of interaction softness in phase separation of active particles}
\author{Monika Sanoria}
\author{Raghunath Chelakkot}
\email{raghu@phy.iitb.ac.in}
\author{Amitabha Nandi}
\email{amitabha@phy.iitb.ac.in}
\affiliation{Department of Physics, Indian Institute of Technology Bombay, Powai, Mumbai, 400076, India.}
\begin{abstract}
Using a minimal model of active Brownian discs, we study the effect of a crucial parameter, namely the softness of the inter-particle repulsion, on motility-induced phase separation. We show that an increase in particle softness reduces the ability of the system to phase-separate and the system exhibit a delayed transition. After phase separation, the system state properties can be explained by a single relevant lengthscale, the effective inter-particle distance. We estimate this lengthscale analytically and use it to rescale the state properties at dense phase for systems with different interaction softness. Using this lengthscale, we provide a scaling relation for the time taken to phase separate which shows a high sensitivity to the interaction softness.
\end{abstract}

\maketitle
\section{Introduction}
The last two decades witnessed a growing interest in the study of \emph{active matter} \cite{Ramaswamy2010,vicsek2012collective,Marchetti2013}, a system microscopically composed of a collection of motile entities that drive the system out-of-equilibrium.  The collective behaviour of such active systems has been studied using particle-based numerical models where individual active agents self-propel along a body-fixed polarity vector. The large-scale collective behaviour crucially depends on the mutual interaction. One well-studied class of models involves interaction between the particles via a pairwise potential, affecting only their positions. For such a system, the interplay between activity and repulsion leads to phase separation, forming a dense phase and a dilute phase. This is commonly known as motility induced phase separation (MIPS)~\cite{Fily2012,Redner2013,Buttinoni2013,vanderLinden2019,wysocki2014cooperative,Cates2015}. Unlike equilibrium phase transition, MIPS occurs in the absence of any attractive interaction \cite{Cates2015,digregorio2018full,caporusso2020motility,stenhammar2013continuum,levis2017active,klamser2018thermodynamic,paliwal2020role,mandal2019motility,caprini2020spontaneous,Das2020,Das2020_pre,Lee_2013,Fily2014,Elgeti_2013,shi2020self,Su_2021,cates,stenhammar2014phase}.  

While the physical principles governing MIPS has been extensively studied before \cite{Bialke_2013, cates, Farage_2015}, a crucial question that remains is how sensitive are they to the softness or deformability of the microscopic entities. The dependence of two-phase coexistence on interaction softness has been observed in passive systems~\cite{Krauth_2015}. To model active colloidal particles that are \emph{practically} non-overlapping, the Weeks-Chandler-Anderson (WCA) form has been used extensively \cite{Weeks1971}. 
Recent studies have also explored the hard particle limit by reducing the overlap beyond WCA and characterized the complete phase behavior \cite{digregorio2018full}. Another interesting limit is to study how an increase in the mutual overlap between the particles affect the collective properties of the system. This limit is relevant from the perspective of microscopic biological entities, such as cells in tissues. Since they are deformable objects, hard potentials may not be a proper representation of the interaction between them. Furthermore, the mechanical properties of the individual entities may vary during the course of development \cite{guo2017cell} which can affect their collective properties. To understand how such a variation in the interaction affects the structural, dynamic, and phase properties, we use a minimal model of active Brownian particles and study their collective behaviour by systematically varying the interaction softness.  We show that the change in phase behaviour with varying softness can be rescaled by an intrinsic lengthscale, which is the effective inter-particle distance. In the following sections, we discuss the model and the numerical techniques, followed by the results and a discussion of the results.

\section{Model}
Our numerical model consists of a collection of $N$ active Brownian particles in 2D that interact with their neighbors repulsively. The repulsive force between the particles at locations $\mathbf{r}_{i}$ and $\mathbf{r}_{j}$ due to excluded volume interaction is obtained from the generalized WCA potential~\cite{zhou2018geometrical},
\beq
\label{Vex}
u(r) = \epsilon^{-1}V_{ex}=
\begin{cases}
4 \left[ \left(\frac{\sigma}{r}\right)^{2\alpha}-\left( \frac{\sigma}{r}\right)^\alpha\right]+1 &;\hspace{0.1cm}\text{$r/\sigma < 2^{1/\alpha}$}\\
0 &;\hspace{0.1cm}\text{$r/\sigma > 2^{1/\alpha}$},
\end{cases}       
\eqn

where $r=|\mathbf{r}_{i}-\mathbf{r}_{j}|$, $\sigma$ is the nominal particle diameter, and the characteristic energy of the system is chosen to be $\beta^{-1} = k_B T$. Here $\alpha$ is the parameter that controls the particle stiffness; lower $\alpha$ denotes soft interaction limit with larger area overlap while upon increasing $\alpha$, the particles become harder and overlap less. The equation of motion of each active particle in non-dimensionalized units is described by the coupled Langevin equations in the overdamped limit (see Appendix~\ref{AppendixA})
\beqr
\label{eq1}
\dot{\bf r}_i &=& -\nabla_i u + \mbox{Pe}~\hat{\mathbf{n}}_i + {\boldsymbol \xi}_i, \\
\dot{\theta}_i&=&{\xi}_i^R.
\label{eq2}
\eqnr
Here, ${\xi}_i$ and ${\xi}_i^R$ are the Gaussian white noise terms with zero mean and satisfy $ \langle \xi_i(t) \xi_j(t') \rangle = 2\delta_{ij} \delta(t-t')$, and $ \langle \xi_i^R(t) \xi_j^R(t') \rangle = 6 \delta_{ij}\delta(t-t')$. The direction of self-propulsion for each particle is $\hat{\mathbf{n}}_i=(\cos{\theta_i},\sin{\theta_i})$. Our minimal model has three relevant control parameters: density $\phi$, the P\'eclet number Pe, which quantifies the activity strength and the  overlap parameter, $\alpha$. Using Eqs.~(\ref{eq1}--\ref{eq2}), we performed Brownian dynamics simulations using an Euler integration step inside a periodic square-box (see Appendix~\ref{AppendixA}). \\~


\section{Results}
We first study the effect of $\alpha$ on density order. 
We benchmark our numerical study to the WCA case ($\alpha = 6$), where the system exhibits a very clear phase separation~\cite{Redner2013}. We aim to quantify any deviation from the known results as $\alpha$ is decreased below 6. To characterize this, we compute the local density $\phi_{\text{loc}}$. For a single uniform phase, the density distribution $P(\phi_{\text{loc}})$ is unimodal while for the phase-separated system $P(\phi_{\text{loc}})$ is bi-modal. We denote the peak values of the dense and dilute phases as $\phi_l$ and $\phi_g$, respectively. 
\begin{figure}[h!]
\centering
\includegraphics[width=0.45\textwidth]{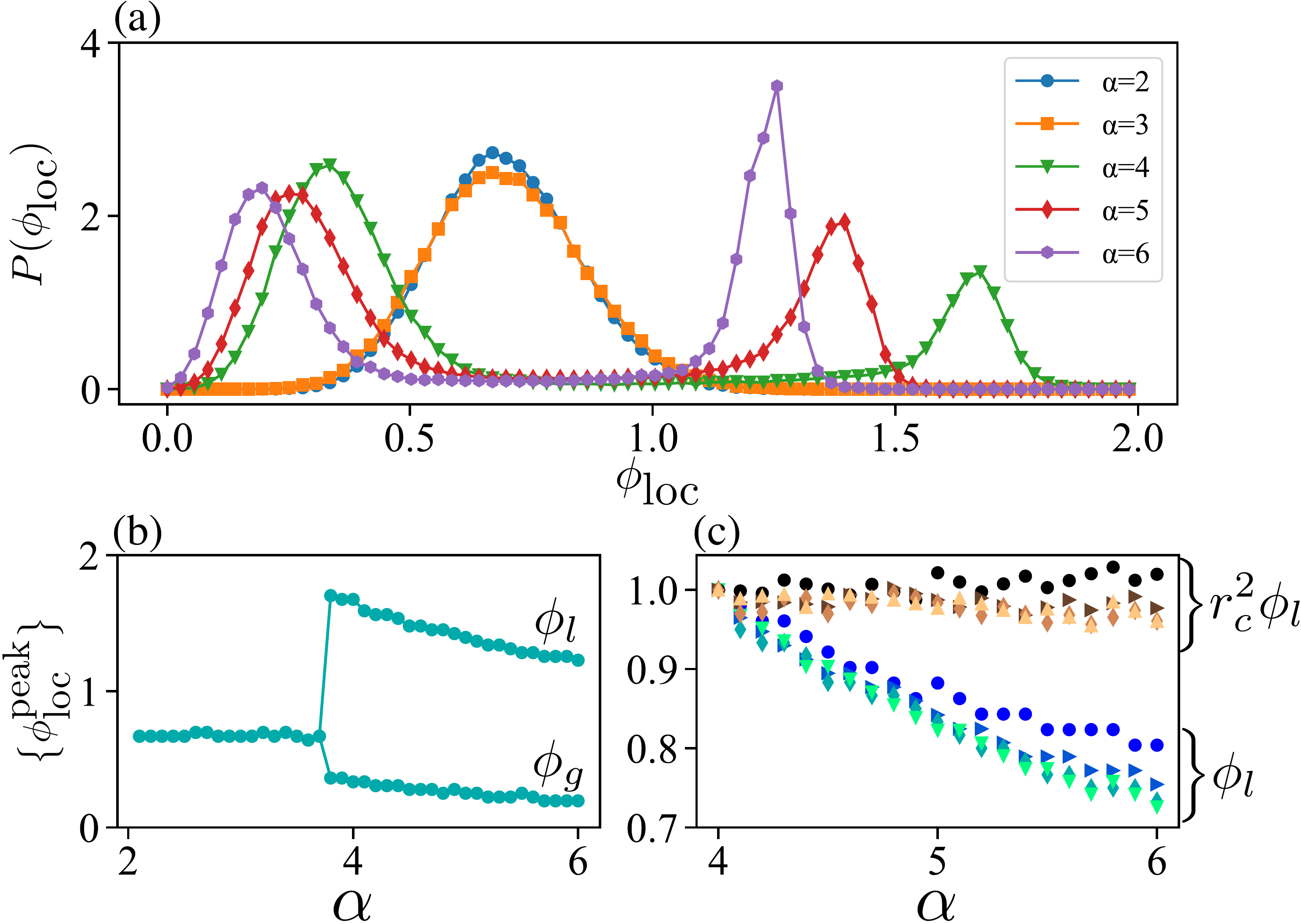} 
\caption{(a) Comparison of the local density distribution $P(\phi_{\text{loc}})$ at $\phi=0.7$ and Pe=150 for different $\alpha$ values. (b) The peak value of the $P(\phi_{\text{loc}})$ distributions observed in (a) are plotted as a function of $\alpha$. The bifurcation indicates the appearance of the dense phase as $\alpha$ is increased. (c) Scaling of the dense phase density $\phi_l$ with $\alpha$ is plotted for Pe=100 ($\bullet$), 130 ($\blacktriangleright$), 150 ($\blacklozenge$), and 170 ($\blacktriangledown$), for $\phi=0.7$. For convenience the $\phi_l$ has been normalized by $\phi_l(\alpha=4)$. For $\phi_l\to r_c^2\phi_l$, the $\alpha$ dependence disappears.}
\label{fig1}
\end{figure}
In Fig.~\ref{fig1}, we compare $P(\phi_{\text{loc}})$ for different $\alpha$ at $\phi=0.7$ and Pe=150. We notice that the phase separation is still observed for $\alpha <6$ (see Fig.~\ref{fig1}(a), for $\alpha=5$ and 4). As we decrease $\alpha$ even further, we do not see a clear phase separation even when the activity is made very high (see Fig.\ref{figs2}, for Pe$=300$ at $\alpha=3$ and $\phi=0.7$ ),  except for a very large density ($\phi \simeq 0.98$ and Pe~$\gtrsim 150$) (see Fig.\ref{figs3}). Such a high density is only accessible for particles with large area overlap and we do not make a detailed comparative study here. Interestingly, we also note that as $\alpha$ is reduced, there is an overall decrease in the area covered by the dense phase (see Fig.\ref{figs1}(b)--(d) in Appendix.~\ref{AppendixA} and Fig.\ref{figs4} in Appendix.~\ref{AppendixB}). To identify phase transition, we plot the corresponding peak values of $P(\phi_{loc})$ as a function of $\alpha$ (see Fig.~\ref{fig1}(b)). We observe, the creation of two distinct phases $\phi_l$ 
and $\phi_g$ for $\alpha \approx 3.7$.  Similar 
disappearance of coexistence with increasing particle 
softness has been observed earlier in 2D equilibrium 
systems as well for an inverse power-law pair 
interaction~\cite{Krauth_2015}.

After phase separation, there is a systematic shift in $\phi_l$, and $\phi_g$ to lower values, as $\alpha$ is increased. In Fig.~\ref{fig1}(c), we have plotted $\phi_l$ as a function of $\alpha$ for different values of Pe at $\phi=0.7$, which clearly shows this monotonic decrease. This is due to the fact that a smaller $\alpha$ leads to a decrease in the average inter-particle distance causing an increase in density. To explain this increase in $\phi_l$ for lower $\alpha$, we use a mean-field description proposed earlier \cite{Bialke_2013}. Within this framework the density dependent effective speed is given by $v(\rho)\equiv\mbox{Pe}-\rho\zeta$, where $\rho$ is the number density and $\zeta$ quantifies the local force-balance in the system and is given by $\zeta=\int_{0}^{\infty}r[-\nabla u(r)]K(r)dr$, where $u(r)$ is the effective pair interaction and $K(r)=\int_{0}^{2\pi}\cos\theta g(r,\theta)d\theta $, where $g(r,\theta)$ is the stationary pair-distribution function. For a single homogeneous phase, and for a given $\alpha$, $\rho(\mathbf{r},t)=\bar{\rho}$ is constant. To estimate the effective cut-off distance $r_c$ between the particles, we consider the scenario where the dense phase has already been formed and we write $\zeta\approx\int_{r_c}^{\infty}r[-\nabla u(r)]K(r)dr$. Furthermore, within the dense phase $v(\bar{\rho})\ll \mbox{Pe}$, we therefore set $v(\bar{\rho})\approx 0$. Note that a systematic estimate of $r_c$ would involve an estimation of $g(r,\theta)$ (and hence $K(r)$) from simulation data, for different $\alpha$. Here we make a simplifying assumption $K(r)\approx K\delta(r-r_c)$, since the dense phase has a crystalline structure where $K(r)$ shows the first peak around $r\simeq r_c$. This leads to:

\beq 
\label{rc}
r_c \approx 4^{1/\alpha} \Bigg{(} 1+\sqrt{1+\frac{2\mbox{Pe}}{\alpha C}} ~\Bigg{)}^{-1/\alpha}.
\eqn

Here $C=\bar{\rho}K$, which in general depends on $\alpha$. Eq.~\ref{rc} can be used to estimate $r_c$ if $C$ is known. As a first approximation we assume $C$ to be a constant \cite{footnote}. To see if this could justify the behavior of $\phi_l$, we estimate $r_c$ using Eq.~\ref{rc} and using $C$ as a fit parameter. We rescale $\phi_l\to r_c^2\phi_l$.  For $C\approx 0.45$, we notice that the dependence on $\alpha$ disappears as $r_c^2\phi_l$ is approximately a constant (see Fig.~\ref{fig1}(c)). 


We also note in Fig.~\ref{fig1}(a)) that with increasing $\alpha$, the distribution of the dense phase about $\phi_l$ becomes narrower, indicating a larger uniformity within the dense phase at higher $\alpha$.  Finally, we note that although for $\alpha \lesssim 3.7$ we do not observe a phase coexistence (see Fig.~\ref{fig1}(b)), the overall phase properties at large Pe are different from a homogeneous phase typically observed at low Pe. Here the system exhibits a complex spatiotemporal dynamics, where we observe a large number of small local clusters that are not stable over time (see Fig.\ref{figs1}(d) in Appendix~\ref{AppendixA} and Movie~1 in \cite{SI} for $\alpha=3$ and Pe = 150). 

\begin{figure}[h!]
\centering
\includegraphics[width=0.5\textwidth]{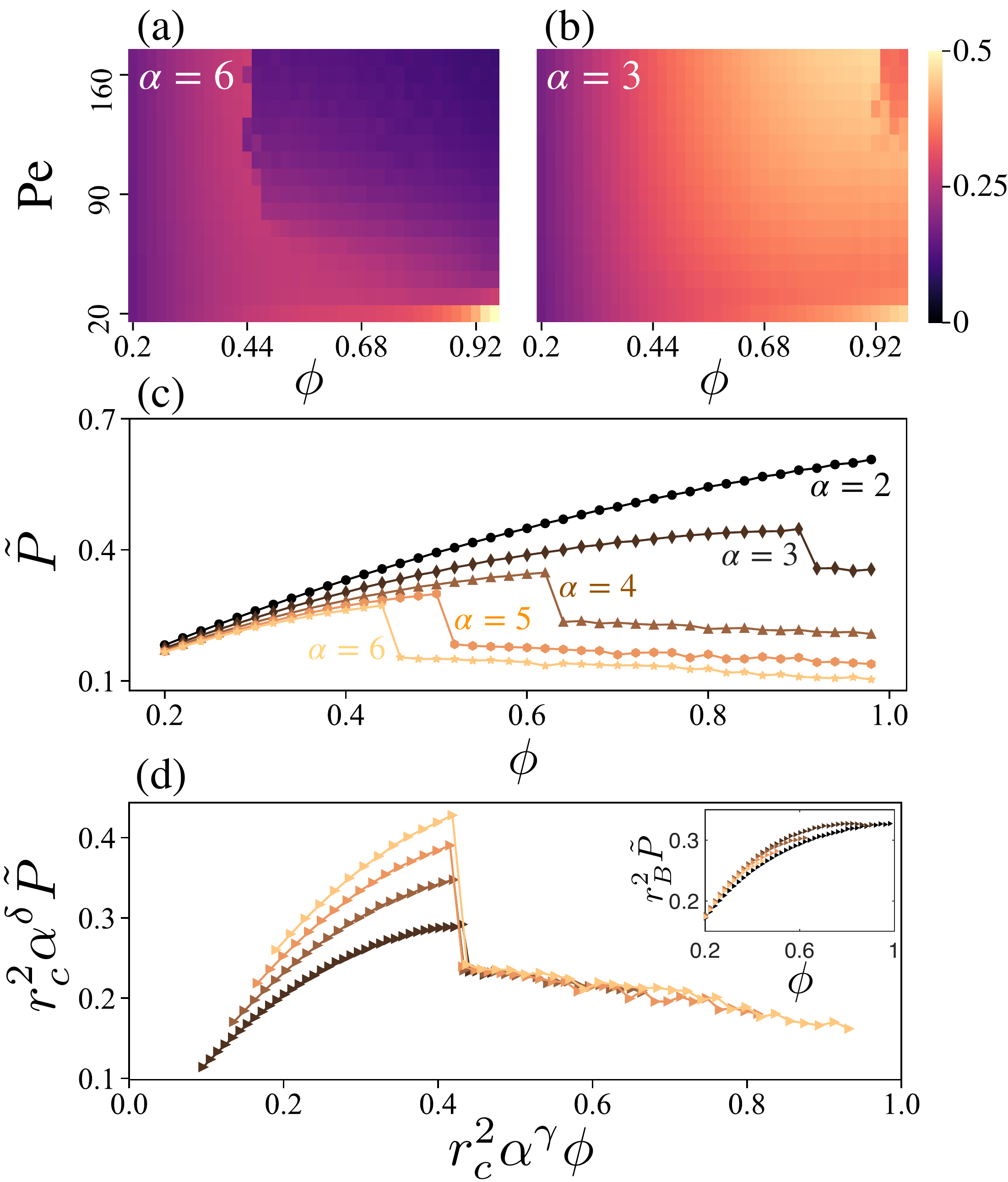}
\caption{Study of the normalized pressure $\tilde{P}=P\phi/P_{id}$. (a) and (b) shows the heat maps of $\tilde{P}$ in the Pe--$\phi$ plane for $\alpha=6$ and $\alpha=3$ respectively. (c) $\tilde{P}$ plotted as a function of $\phi$ at Pe=150 for $\alpha = 2 \to 6$. (d) Study of scaling behavior for the phase separating cases shown in (c). The $\phi$-axis is scaled by,~\ie~$r_c^2\alpha^\gamma$, with $\gamma\approx0.17$, while the pressure-axis is scaled with a factor $r_c^2\alpha^\delta$ ($\delta\approx 0.45$). The inset show the scaling of $\tilde{P}$ before phase separation, using the Boltzmann diameter $r_B$.
}
\label{fig3}
\end{figure}

The appearance of dense phase and its dependence on $\alpha$ can be understood by quantifying the pressure in  presence of activity~\cite{Brady2014,Takatori2014,SolonFily2014,SolonPRL,Winkler2015,takatori2016forces}.
We calculate the normalized pressure $\tilde{P}=P\phi/P_{id}$, where $P$ is the total pressure, calculated by combining the contributions from both thermal fluctuations and activity \cite{Winkler2015} for the our system (See Eq.~\ref{Pex} in Appendix~\ref{AppendixC}), and $P_{id}=\dfrac{N}{A}(1+\dfrac{\text{Pe}^2}{6})$ \cite{Winkler2015}. Here $A$ is the total area in non-dimensional units. In Fig.~\ref{fig3} we show $\tilde{P}$ for $\alpha=6$ (Fig.~\ref{fig3}(a)) and 3 (Fig.~\ref{fig3}(b)) as a function of $\phi$ and Pe. At $\alpha = 6$, we observe a drop in the value of $\tilde{P}$ in the $\phi$-Pe parameter space where the system phase separates. For $\alpha=3$, the phase separation is shifted to a large value of $\phi$ and Pe. This is visible in Fig.~\ref{fig3}(b) --- we observe a small region in $\phi$-Pe parameter space where $\tilde{P}$ drops to a smaller value.  For a given Pe, the $\alpha$ dependence of the normalized pressure $\tilde{P}$ is summarized in Fig~\ref{fig3}(c). We see how, with decreasing $\alpha$, the critical density $\phi_{cr}$, shift to higher densities until it disappears. This systematic shift of the  critical point, can be explained as the following: we note that for all $\phi>\phi_{cr}$, $\phi=(N_l+N_g)\pi/A$, where $N_l$ and $N_g$ are the number of particles in the dense and dilute phases. Based on the numerical evidence (see Fig.\ref{figs5} in Appendix.~\ref{AppendixD}), we assume $\phi_g\approx f_{\alpha}\phi_l$, where $f_{\alpha}<1$ is a scale-factor with a weak dependence on $\alpha$. This leads to $\phi\approx\phi_l h(\alpha)$, where $h(\alpha)$ is an unknown function that explicitly depends on the area occupied by the dilute phase (see Appendix~\ref{AppendixD}). Making a simplifying assumption that $h(\alpha)$ have a power-law dependence on $\alpha$, we scale the $\phi$-axis,~\ie~$\phi\to r_c^2\alpha^\gamma\phi$ for different $\alpha$ values where phase separation has been observed. Using the $r_c$ value calculated previously using Eq.~\ref{rc}, and for $\gamma\approx 0.17$, we are able to collapse the transition points along the $\phi$-axis indicating that the critical density $\phi_{cr}$ also follows the same scaling (see Fig.~\ref{fig3}(d)). The small value of the exponent $\gamma$ is consistent with our prediction that $h(\alpha)$ is weakly dependent on $\alpha$ (see Appendix~\ref{AppendixD}). The bulk pressure in the dense phase also follows a similar scaling, \ie~$\tilde{P}\to r_c^2\alpha^\delta\tilde{P}$ with $\delta \approx 0.45$ (Fig.~\ref{fig3}(d)), implying the qualitatively same $\alpha$ dependence as seen for the case of $\phi$. Note that before phase separation, $\tilde{P}$ can be rescaled by the Boltzmann diameter $r_B=2^{1/\alpha}(1+\sqrt{g_{\alpha,\phi}})^{-1/\alpha}$ derived from equilibrium considerations \cite{zhou2018geometrical}. Here $g_{\alpha,\phi}$ is a collision parameter that depends on both the softness and the density. We estimate $g_{\alpha,\phi}$ from \cite{zhou2018geometrical} (see Appendix~\ref{AppendixE}) and rescale $\tilde{P}\to r_B^2\tilde{P}$. As shown in the inset of Fig.~\ref{fig3}(d), the curves collapse reasonably well, especially at low $\phi$ values, as expected.

Having looked into the dependence of density ordering and the bulk pressure on the overlap parameter $\alpha$, we now study how the structural properties of the system is affected by changing this parameter. For this purpose we study the global orientational order, $\psi_6=\bigg\langle \left| \frac{1}{N}  \sum \limits_{i=1}^{N} \phi_{6i} \right|\bigg \rangle$, where $\phi_{6i}=\frac{1}{N_b} \sum \limits_{j \in N_b} e^{6 i \theta_{ij}}$, calculated at the location of every particle  $i$. Here $N_\text{b}$ is the total number of neighbours of the $i^{\mbox{th}}$ particle obtained using Voronoi tessellation, \(\theta_{ij}\) is the angle between the vector $\mathbf{r}_{ij}$ and the reference axis.
 
\begin{figure}[h!]
\centering
\includegraphics[width=0.49\textwidth]{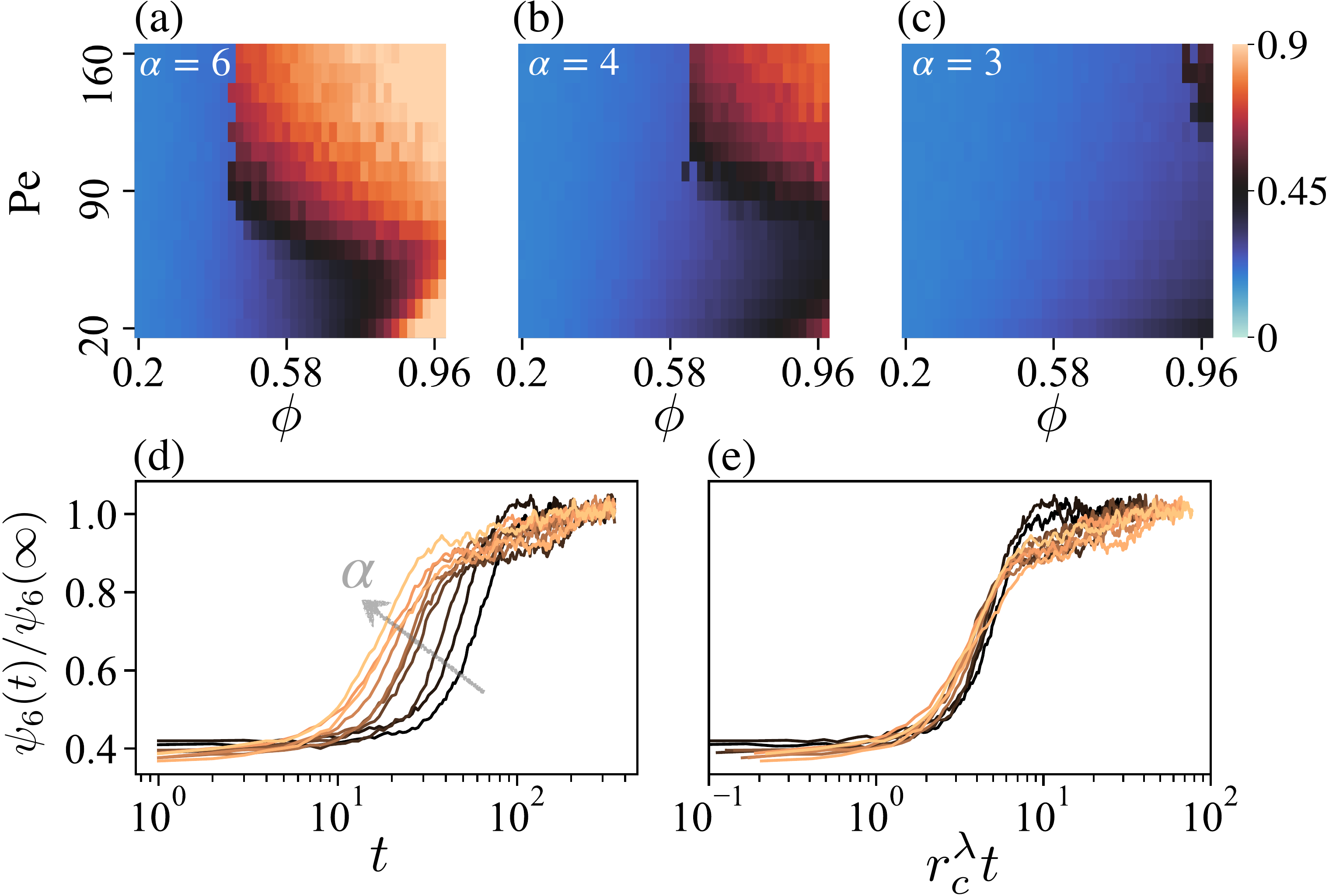}
\caption{Quantification of global orientational order. (a)--(c) The global orientational order parameter $\psi_6$ is shown as a colormap in the $\phi$--Pe plane. Dynamics of phase-separation as a function of $\alpha$. (d) $\psi_6(t)/\psi_6(\infty)$ ($\psi_6(\infty)$ is the steady state value) plotted as a function of time for $\alpha=4\to6$ with $\Delta\alpha=0.2$. Here Pe=150 and $\phi=0.7$. (e) Demonstration of temporal scaling. Introducing a scaling $t\to r_c^{\lambda}t$, leads to a collapse of all the different plots for $\lambda\approx8.5$.}
\label{fig2}
\end{figure}

In Fig.~\ref{fig2}(a)--(c) we compare the $\psi_6$ heatmaps in the $\phi$--Pe plane for $\alpha=6,4$, and 3. We notice that the phase separated domain, indicated by a large $\psi_6$ ($\gtrsim 0.4$) value, gradually decreases in size in the $\phi$-Pe plane as $\alpha$ is varied from $6\to3$. Furthermore, the phase boundary (defined by $\psi_6\approx 0.4$) also gradually shifts towards higher values of Pe and $\phi$, as $\alpha$ is decreased. There is also an increase in the sixfold symmetry within the dense phase for larger $\alpha$, due to an increase in the cluster size (see Fig.\ref{figs4} in Appendix.~\ref{AppendixB}). This behaviour is consistent with our observation of the local density distributions shown in Fig.~\ref{fig1} for $\phi=0.7$. Using the same scaling relation for $\phi_{cr}$ in the pressure analysis (Fig~\ref{fig3}(d)), we rescaled  $\phi\to r_c^2 \alpha^{\gamma}\phi$, which leads to a collapse of $\phi_{cr}$ for different $\alpha$ (See Fig.\ref{figs6}(b) in Appendix.~\ref{AppendixF}). We also note that $\psi_6 \sim r_c$ after phase separation  (See Fig.\ref{figs6}(b)). 

We also note for $\alpha=6$, at low Pe and large $\phi$, there is an enhanced global order (see Fig.~\ref{fig2}(a)). Such behavior has also been observed recently in systems of active Brownian particles, even with lesser overlap \cite{digregorio2018full}. This feature is not seen for $\alpha=4$ and 3 (Fig.~\ref{fig2}(e)--(f)) since the larger overlap between the particles destroys the hexatic order at such high densities.

To understand how the emergence of phase ordering changes with varying $\alpha$, we plot $\psi_6(t)$ versus time starting from a homogeneous phase for different values of $\alpha$. While a more detailed approach to address this question will involve the study of coarsening dynamics, the $\psi_6$ dynamics can also be used to investigate this transition. In Fig.~\ref{fig2}(d), we plot $\dfrac{\psi_6(t)}{\psi_6(\infty)}$ as a function of time, where $\psi_6(\infty)$ is the steady-state value after the system has already phase-separated. We notice that with a decrease in $\alpha$, the system takes longer to phase-separate. To quantify this dependence on particle softness, we apply a rescaling of the time axis, $t\to r_c^{\lambda}t$. We use $r_c$ instead of $\alpha$ for rescaling (see Fig.\ref{figs7}(b) in Appendix.~\ref{AppendixF}), since it is an intrinsic length-scale, which can be obtained in general for any form of the potential used. This rescaling leads to a data collapse for $\lambda \approx 8.5$ (see Fig.~\ref{fig2}(e)). The large value of $\lambda$ indicative of sensitive dependence of the transition time on $\alpha$. This dependence can also be of an exponential form (see Fig.\ref{figs7}(c) in Appendix.~\ref{AppendixF}).

\begin{figure}[h!]
\centering
\includegraphics[width=0.45\textwidth]{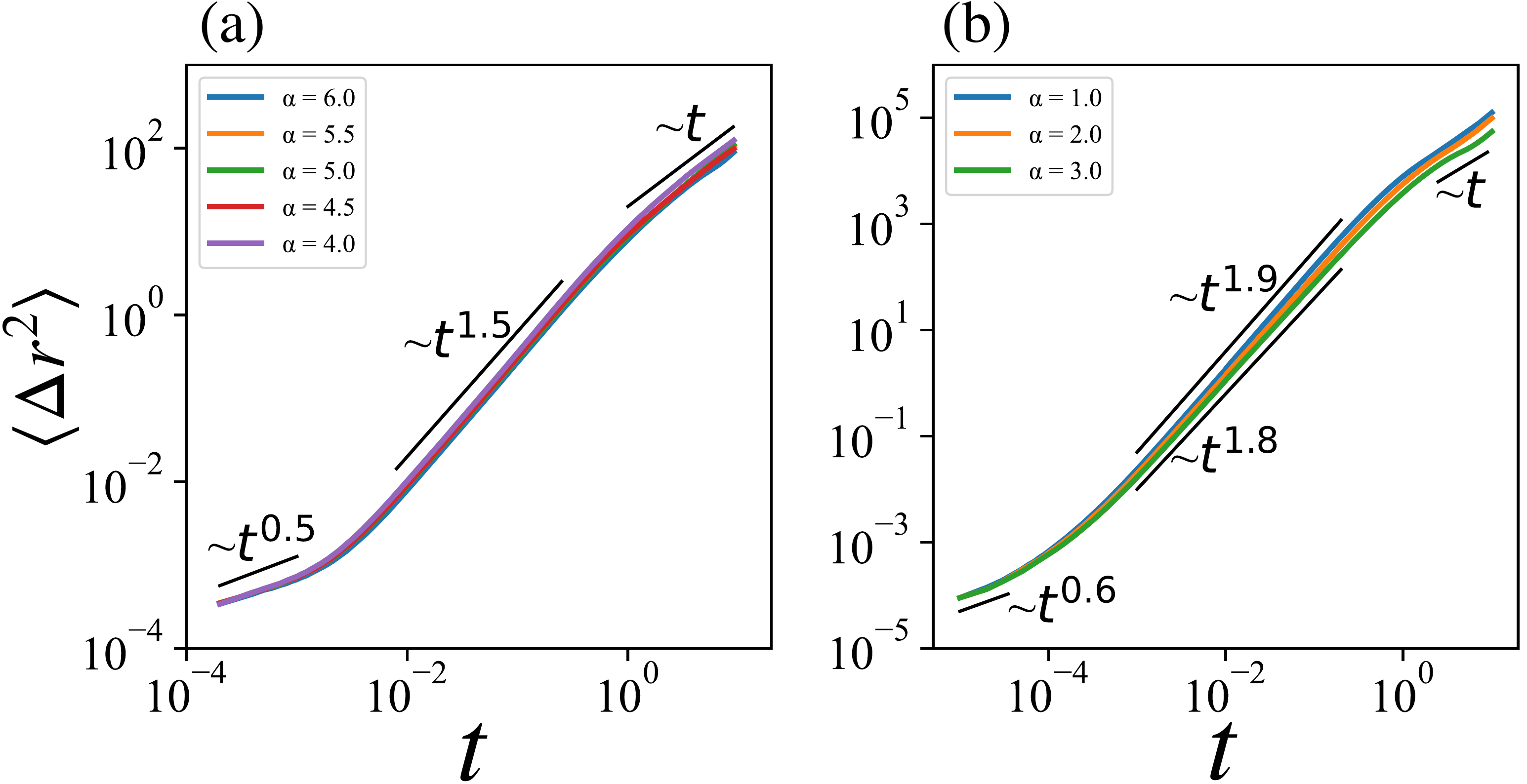}
\caption{(a) Mean-square displacement $\la{\D \mathbf{r}^2(t)}\ri$ of a particle inside the dense phase for different $\alpha$. (b) $\la{\D \mathbf{r}^2(t)}\ri$ plotted for $\alpha=3$ in the case when the system does not phase separate to form a single large cluster. Here Pe=150 and $\phi=0.7$.}
\label{fig4}
\end{figure}

Our study so far has revealed the dependence of the static quantities, such as $\phi_{cr}$, $\phi_l$, $\tilde{P}$, and $\psi_6$ on $\alpha$, in a phase-separated system, through a system lengthscale, $r_c$, defined for the dense-phase. Also, it shows that the time required for the system to phase separate is sensitive to $\alpha$. A natural question arises whether the transport properties within the dense-phase also vary with $\alpha$. To study this, we compute the mean-square displacement (MSD) of one or a few tagged particles inside the dense phase. In Fig.~\ref{fig4}, we show the MSD of a tag-particle, inside the dense phase, as a function of lag-time ($t$) for $4.0 \leq \alpha \leq 6$, Pe=150, and $\phi=0.7$. We note that the MSD for $\alpha=6$ shows good agreement with the reported results~\cite{Redner2013}. Interestingly, the MSD behaviour did not show any noticeable change as $\alpha$ is varied. This indicates that once the system phase-separates, unlike the static quantities, the particle dynamics is almost independent of $\alpha$.  This observation is further strengthened by the MSD behaviour for $\alpha =3$ and 2 at $\phi=0.7$ and Pe=150, where there is no single large cluster but the formation of many small local dynamic clusters. In this case, for the same values of $\phi$ and Pe, we notice a deviation in the MSD behaviour (see Fig.~\ref{fig4}(b)), qualitatively different from the previous observations in both the dilute phase and dense phase (see Fig.\ref{figs8} in Appendix.~\ref{AppendixG}), indicating a different collective dynamics which needs to be studied in detail.

\section{Conclusion}
Our study reveals that an increase in interaction softness has an inhibiting effect on phase separation and strongly influences the collective phase ordering: the critical density above which the system phase separates shifts to higher values, while the waiting time for observing a phase separation, above the critical density, also increases. The phase separation in the system is manifested by an intrinsic lengthscale, which is the effective inter-particle distance $r_c$. The changes in state properties with interaction softness in the dense phase can be explained using $r_c$. Using a mean-field theory \cite{Bialke_2013}, we make an approximate estimate of $r_c$ which explain the $\alpha$ dependence of the dense phase reasonably well. Using a relation between the dense and dilute phase densities obtained numerically, we show that the critical density at the onset of phase ordering and the corresponding bulk pressure can be scaled using the same $r_c$ with power-law corrections. Before the phase separation, the rescaling can be achieved reasonably by the Boltzmann diameter, $r_B$. To characterize the temporal efficiency as a function of $\alpha$, we also studied the time evolution of the global hexatic order for different $\alpha$. We found that an increase in particle softness (or decrease in $\alpha$), significantly slows down the approach to transition and scales as $\sim r_c^{8.5}$. However, the transport properties within the dense phase show no variation with $\alpha$. This indicates that once the system phase separates, the dynamic behaviour inside the dense phase is weakly dependant on the microscopic interaction parameters. Note that the disappearance of phase separation has been observed in inertial systems, with an increase in Pe to $\epsilon$ ratio~\cite{Suma_2014, mandal2019}. It would be interesting to explore, how an increase in interaction softness modify this behaviour.

The dependence of phase separation on interaction softness has relevance beyond simple models of active matter. The mutual interactions in many active biological systems are not hard-core in nature and may vary across systems. Our study thus may provide a physical basis for phase ordering in such systems. This behaviour need not be specific to the particular form of interaction we have used, and we expect to observe similar behaviour with other forms of short-ranged repulsive interactions.

\section*{Acknowledgements}
The authors thank Debasish Chaudhuri, Dibyendu Das, and Mithun K. Mitra for insightful discussions. MS thank CSIR, India for financial support. AN and RC acknowledges Industrial Research and Consultancy Centre (IRCC) at IIT Bombay, India, and Science and Engineering Research Board (SERB), India (Project No. ECR/2016/001967, ECR/2017/000744, and SB/S2/RJN-051/2015) for financial support. We thank IIT Bombay HPC facility (Spacetime2).



\appendix
\section{Simulation Method} 
\label{AppendixA}
Our numerical model consists of a collection of active Brownian particles in 2D that interact with its neighbors repulsively. The repulsive force ${\bf F}_{ex}(\mathbf{r})$ between the particles at locations $\mathbf{r}_{i}$ and $\mathbf{r}_{j}$ due to excluded volume interaction is obtained from the generalized WCA potential \cite{zhou2018geometrical} $V_{ex}(r)$ given in Eq.~\ref{Vex}.

\begin{figure}[h!]
\centering
\includegraphics[width=0.35\textwidth]{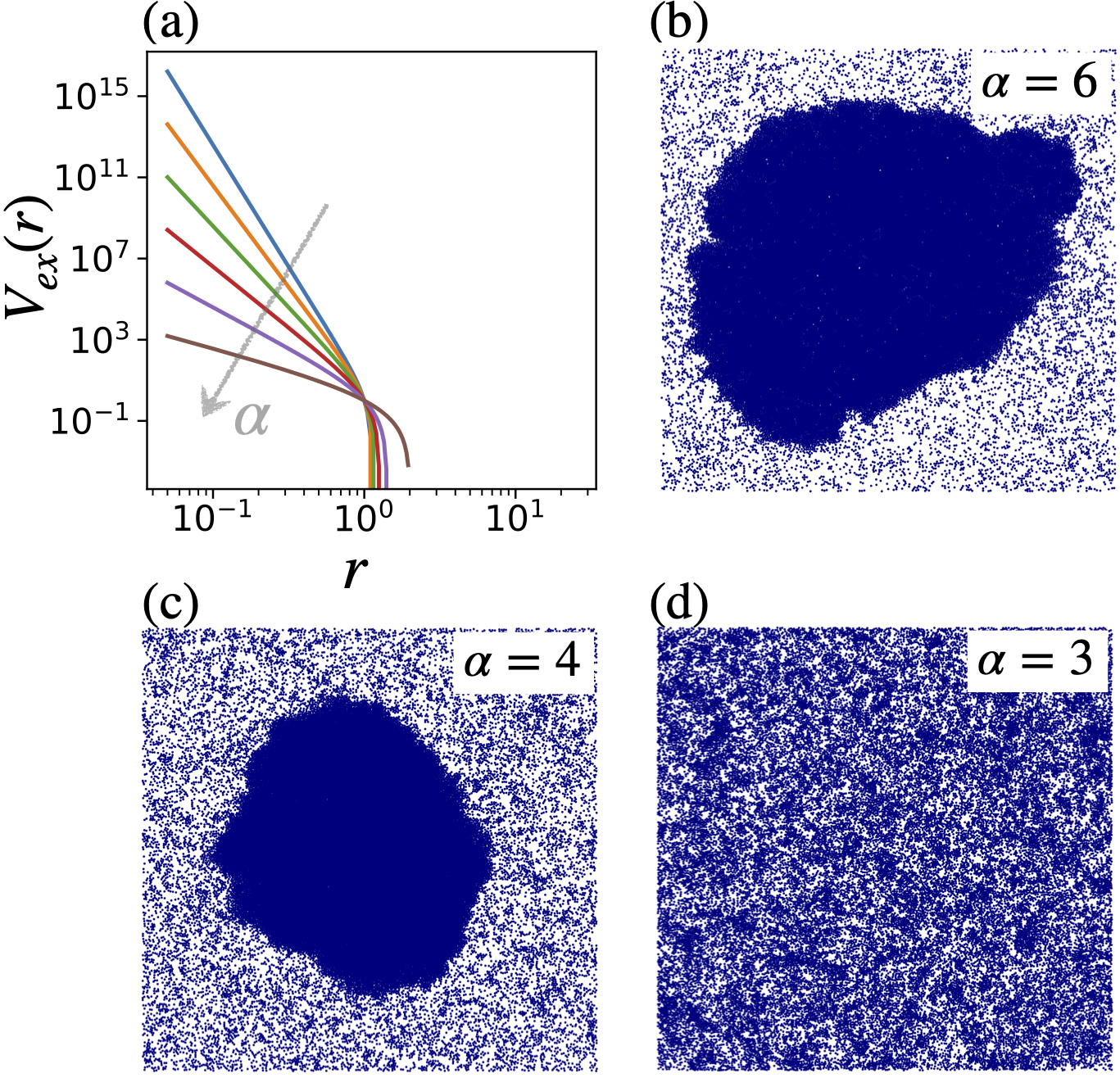}
\caption{Effect of the overlap parameter $\alpha$ on the phase behavior.  (a) Interaction potential $V_{ex}(\mathbf{r}_{ij})$ plotted against ${r}_{ij}$ for different values of $\alpha$ varied from $6\to1$. (b)--(d) shows the steady state snapshots indicating the phase behavior for a system of $N = 40,000$ particles, at $\phi = 0.7$, Pe=150.}
\label{figs1}
\end{figure}

Each particle is represented by a position vector $\mathbf{r}_i$ and moves with self-propelled constant speed $v_{p}$ along a direction $\mathbf{n}_i$, where $\mathbf{n}_i=(\cos{\theta_i},\sin{\theta_i})$. The particle direction $\mathbf{n}_i$ are evolved randomly in time and equation of motion of active particle is described by the coupled and  over-damped Langevin equations
\beqr
\label{eq1_nondim}
\dot{\bf r}_i &=& v_p \hat{\mathbf{n}}_i + \mu{\bf F}_{ex}({\mathbf{r}})  + {\boldsymbol \xi}_i, \\
\dot{\theta}_i&=&{\xi}_i^R.
\label{eq2_nondim}
\eqnr

Eqs.~(\ref{eq1_nondim}--\ref{eq2_nondim}) are non-dimensionalized to obtain Eqs.~(\ref{eq1}--\ref{eq2}) using the particle diameter $\sigma$, thermal energy $\beta^{-1}=k_B$T and relaxation time $\tau = \frac{\sigma^2}{D_t}$. The characteristic energy of the system is chosen to be $\epsilon=\beta^{-1}$. Here, $v_p$ is the active self propulsion speed of each particle and  $\mu=\beta D_t$ is the mobility parameter, $D_t$ is translational diffusion coefficient, and $D_r = 3D_t/\sigma^2$, is the rotational diffusion coefficient. ${\boldsymbol \xi}_i$ and ${\xi}_i^R$ are the Gaussian white noise terms with mean zero and variance unity and satisfies $ \langle {\boldsymbol \xi}_i(t) {\boldsymbol \xi}_j(t') \rangle = 2D_t\delta_{ij} \delta(t-t')$, and $ \langle \xi_i^R(t) \xi_j^R(t') \rangle = 2D_r \delta(t-t')$.

The effect of activity is studied by varying the P\'eclet number (Pe), a non-dimensional parameter defined as $\mbox{Pe} = v_p \frac{\tau}{\sigma}$. Our minimal model has three relevant control parameters: density $\phi$, activity strength Pe, and interaction parameter $\alpha$. To characterize the system, these parameters are varied such that, $\phi\in[0.1,0.98]$, Pe~$\in[20, 180]$ and $\alpha\in[2,6]$.

We performed Brownian dynamics simulations using Euler integration step inside a periodic square-box for $N = 40000$ to $112000$ particles and with Brownian time step $\Delta t=10^{-5} \tau$ upto $t=100\tau$ . The translational diffusion coefficient is chosen to be $D_t=1$ throughout.  

\section{Phase behavior for the case of $\alpha<6$}
\label{AppendixB}
\begin{figure}[h!]
\centering
\includegraphics[width=0.35\textwidth]{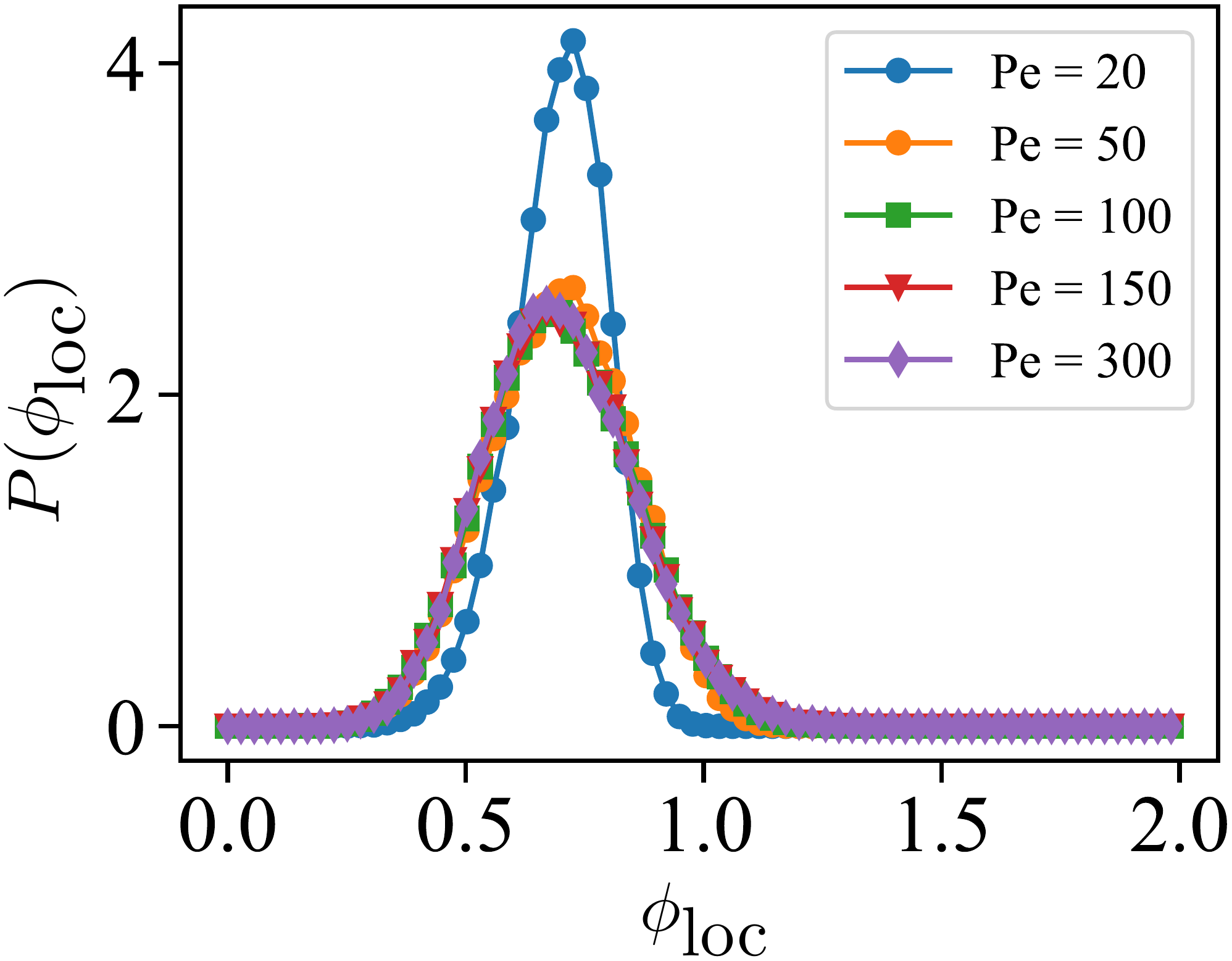}
\caption{ Local density distribution $P(\phi_{\text{loc}})$ with varying Pe at $\phi=0.7$ and $\alpha=3$. We notice that the system does not phase separate even at large value of activity (Pe=300).}
\label{figs2}
\end{figure}

\begin{figure}[h!]
\centering
\includegraphics[width=0.45\textwidth]{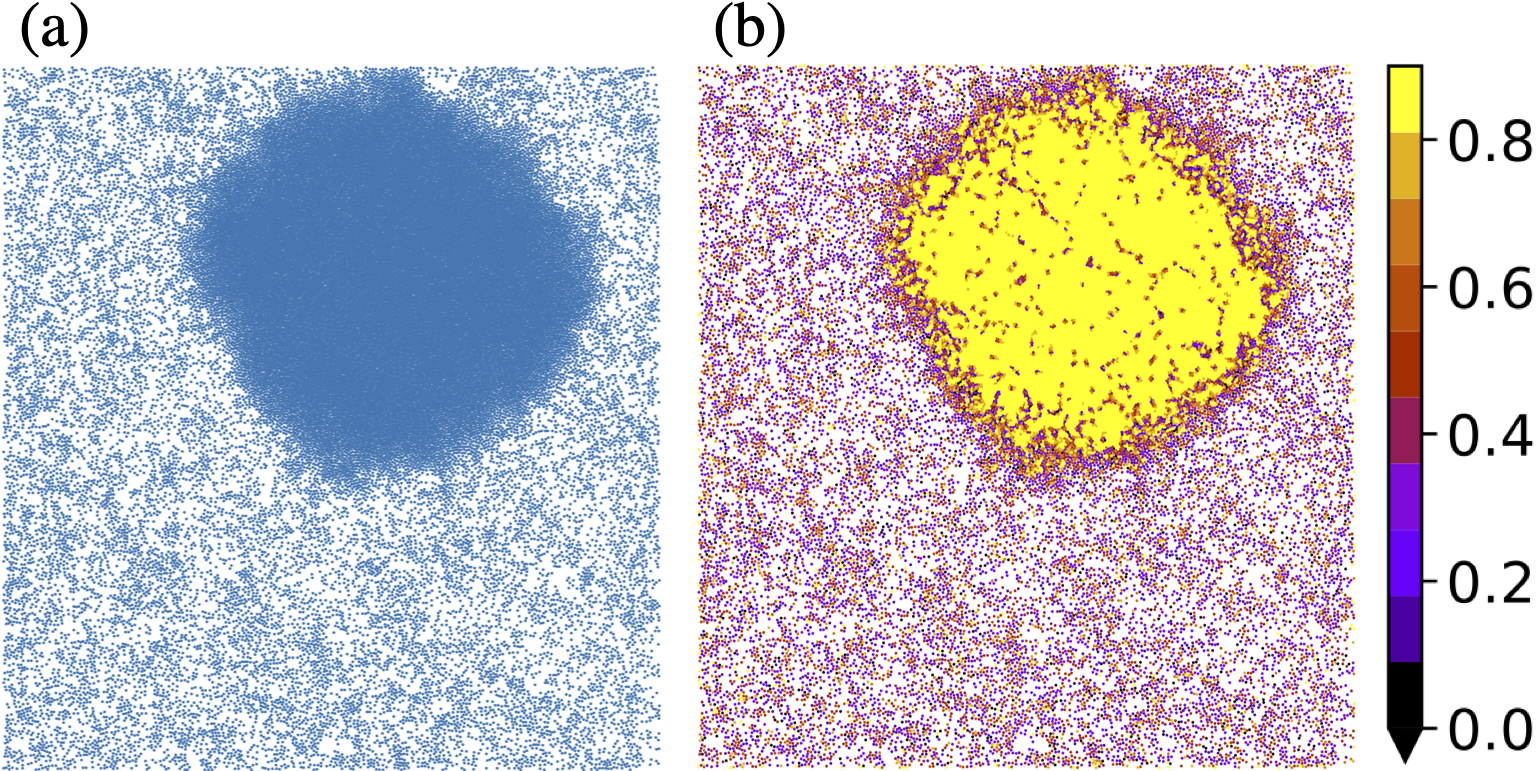}
\caption{(a) Phase separation observed for $\alpha=3$ at very large densities ($\phi=0.98$). He we set Pe=170 and the simulation was run for a system of $N = 40,000$ particles. (b) shows the local orientational order $\phi_{6i}$ as a color coded value superimposed on particle configurations.}
\label{figs3}
\end{figure}
\begin{figure}[h!]
\centering
\includegraphics[width=0.45\textwidth]{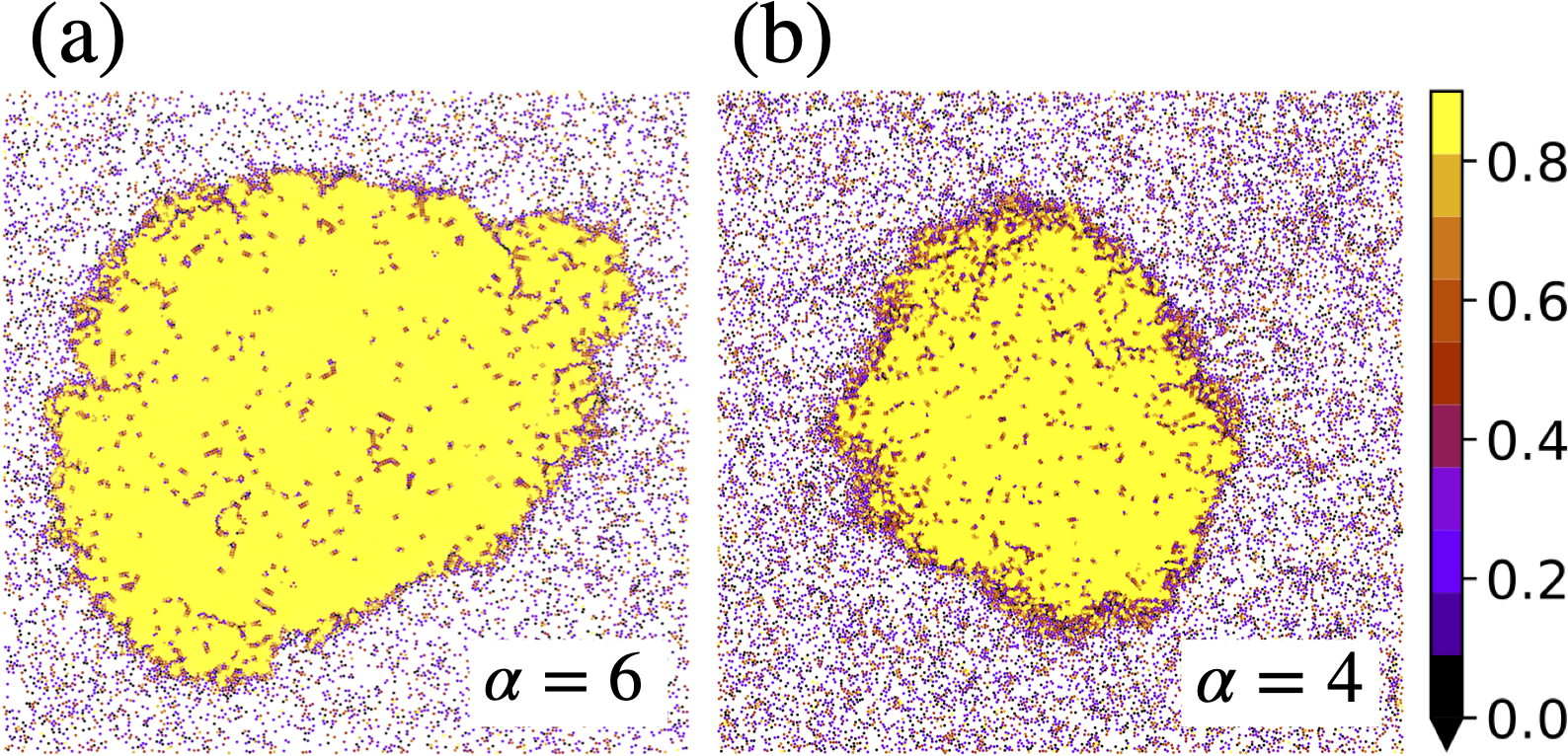}
\caption{Local orientational order $\phi_{6i}$ as a color coded value superimposed on particle configurations in the $x-y$ plane for different $\alpha=6$ and 4 respectively. Here we set $\phi=0.7$ and Pe=150. For $\alpha=6$, we notice that there are few defect points of mainly 5-7 pairs present within the dense region (see (a)). Similar defect points are also observed in the dense phase for $\alpha=4$ but the number is higher. This is possibly caused by the large overlap between the particles in this case causing deviations from the six-fold symmetry locally.}
\label{figs4}
\end{figure}

\section{Active pressure for 2D active system}
\label{AppendixC}
Total pressure of a 2D active system with area $A$ and total number of particles $N$, can be calculated using the following expression \cite{Winkler2015}
\beq
\label{Pex}
P=\frac{1}{A}\bigg\{N + \text{Pe}\frac{Nv}{6} + \frac{1}{4} \sum_{i=1}^{N} \sum_{j=1}^{N} \sum_{n}\langle \mathbf{ F}^n_{ij} \cdot(\mathbf{ r}_i-\mathbf{ r}_j -\mathbf{ R}_{ij}^n ) \rangle \bigg\}.
\eqn
Here the first term on the right hand side is the thermal contribution and the second term is the swim pressure. Here, $v = \frac{1}{N}     \sum_{i=1}^{N} \langle  \mathbf{\dot r}_i \cdot  \hat{\mathbf{n}}_i \rangle$ is defined as the mean speed of the active particle. The last term is contribution due to inter-particle interaction. The last two terms in Eq.~\ref{Pex} is called the total active pressure.\\

 \section{Critical Density}
\label{AppendixD}
  Let $\phi_{cr}$ be the critical density at which phase-separation occurs. We assume there are $N_l$ number of particles in the dense (liquid) phase and $N_g$ in the dilute (gas) phase. Let $A_l$ and $A_g$ be the area of the dense and dilute phases respectively. Therefore, $\phi_{cr}$ can be written as
\beq
\label{phi_cr}
\phi_{cr} = \frac{N_l + N_g}{A}=\phi_l \frac{A_l}{A} + \phi_g \frac{A_g}{A}.
\eqn
\begin{figure}[h!]
\centering
\includegraphics[width=0.46\textwidth]{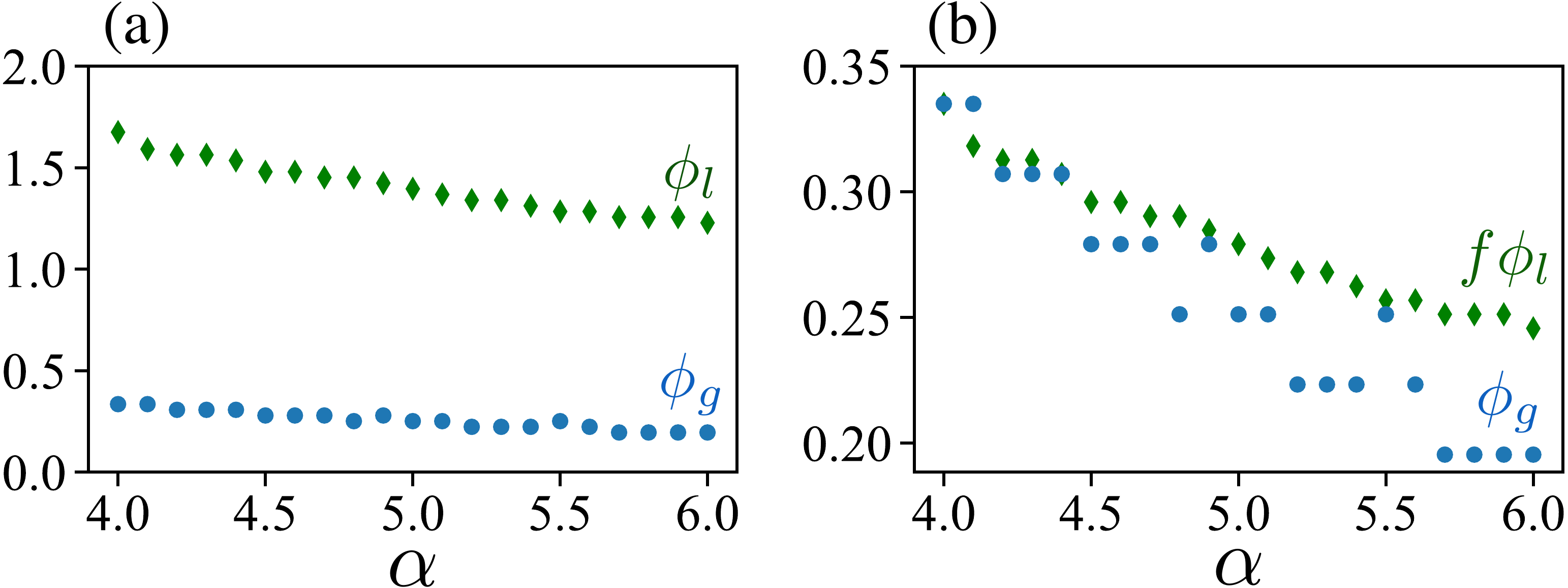}
\caption{(a) Values of $\phi_l$ ($\blacklozenge$) and $\phi_g$ ($\bullet$) plotted as a function of $\alpha$ at $\phi=0.7$ and Pe=150. (b) To compare the trend we multiply $\phi_l$ by a factor of $f=0.2$. 
}
\label{figs5}
\end{figure}
By comparing the values of $\phi_l$ and $\phi_g$ obtained numerically (see Fig.~\ref{figs5}), we note that both shows a similar dependence on $\alpha$ and we can assume $\phi_g \approx f_{\alpha} \phi_l$. Here $f_{\alpha}$ is weakly dependent on $\alpha$ with magnitude $<1$. Substituting this in Eq.~\ref{phi_cr}, we obtain
\beq
\phi_{cr} \approx \phi_l \Big[ 1 -\big(1 - f_{\alpha}\big)\ \frac{A_g}{A}  \Big] \approx \phi_l h\big( \alpha \big)
\eqn
Here $\big( 1 - f_{\alpha}\big)<1$ and weakly dependent on $\alpha$ and $\frac{A_g}{A}$ is also a function of $\alpha$ with magnitude less than unity. Therefore the quantity $h(\alpha)$ is expected to have a weak dependence on $\alpha$ as well.

\section{Scaling using the Boltzmann diameter}
\label{AppendixE}
Taking the collision energy to be of the order of $K_BT$, one can derive the typical interparticle distance, namely the Boltzmann parameter given by \cite{zhou2018geometrical}
\beq
r_B=\Bigg{(}\frac{2}{1+\sqrt{g_{\alpha,\phi}}}\Bigg{)}^{1/\alpha}
\eqn
Here $g_{\alpha,\phi}$ is a collision parameter and depends on both the $\alpha$ and $\phi$. {We assume a functional form for the collision parameter $g_{\alpha,\phi} \sim \alpha^{a_1}\phi^{a_2}$ where $a_1$ and $a_2$ are constant exponents. Using the simulation data reported in Ref.~\cite{zhou2018geometrical} (Table-I)), we obtain $a_1\approx-0.74$, and $a_2\approx 1.12$. The estimated values of $r_B$ using a form of $g_{\alpha,\phi}\approx12.54~\alpha^{-0.74}\phi^{1.12}$ leads to a reasonably good scaling of $\tilde{P}$ (shown in the inset of Fig.~2(d)) before phase-separation.}

\section{Hexatic order parameter scaling}
\label{AppendixF}

\begin{figure}[h!]
\centering
\includegraphics[width=0.45\textwidth]{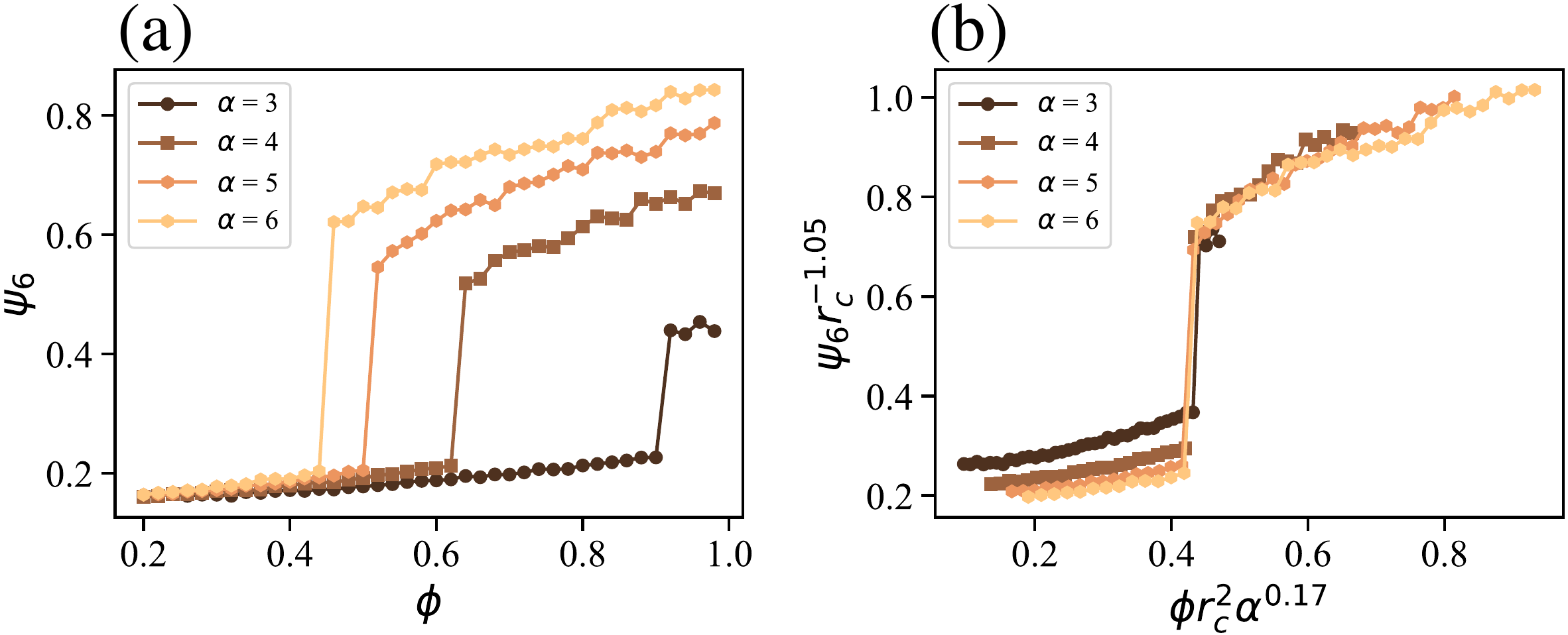}
\caption{(a) $\psi_6$ plotted as a function of $\phi$ at Pe=150 for $\alpha=3\to6$. (b) Study of scaling behavior. For the phase separating cases shown in (a), the $\phi$-axis is scaled as $\phi\to r_c^2\alpha^{0.17}\phi$, while the $\psi_6$ is scaled as $\psi_6\to r_c^{-1.05}\psi_6$. 
}
\label{figs6}
\end{figure}

Similar to the $\tilde{P}$ versus $\phi$ plot shown in Fig.~\ref{fig2}(c), $\psi_6$ also shows a sharp change in value at a critical density value $\phi_{cr}$ (see Fig.~\ref{figs6}(a)). Using the same scaling as used in Fig.~\ref{fig2}(d) (\ie $\phi\to r_c^2\alpha^{0.17}\phi$), we scale the $\phi$-axis. The result is shown in Fig.~\ref{figs6}(b). We also scaled the $\psi_6\to r_c^{-1.05}\psi_6$, which leads to a collapse of the $\psi_6$ value after phase separation.

\begin{figure}[h!]
\centering
\includegraphics[width=0.48\textwidth]{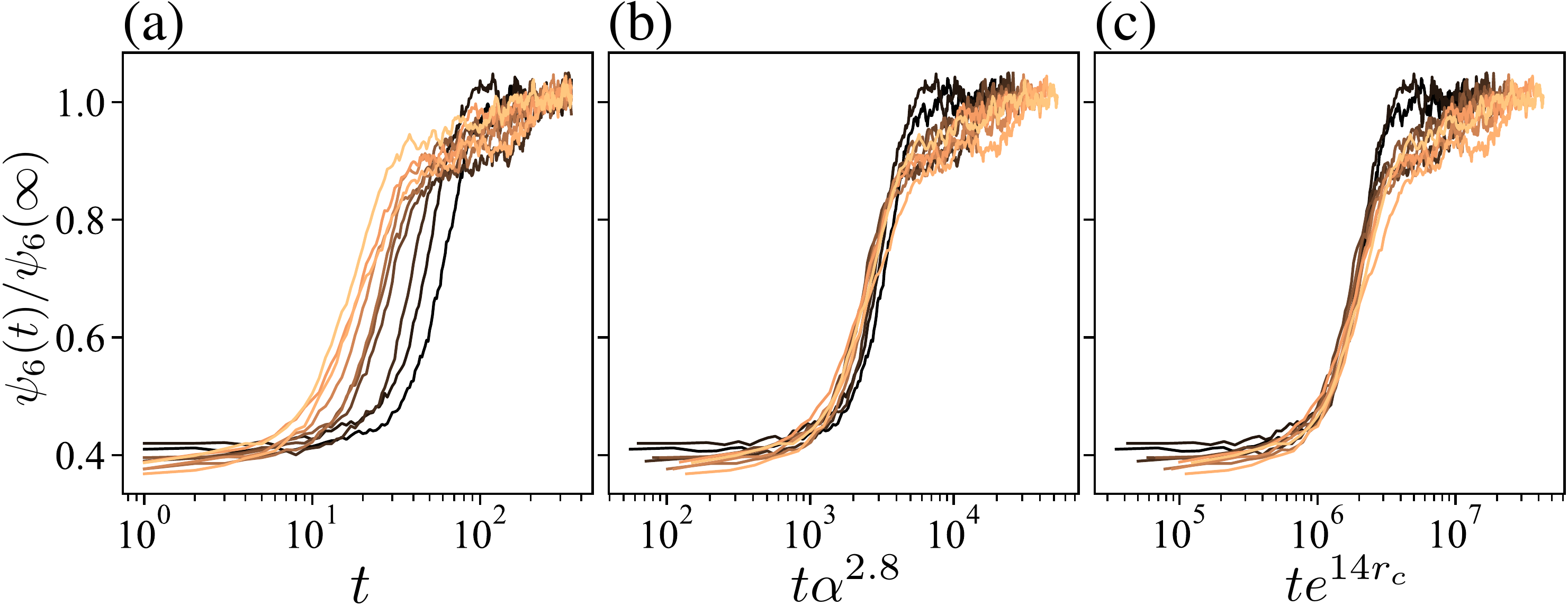}
\caption{(a) ${\psi_6(t)}/{\psi_6(\infty)}$ plotted as a function of time, where $\psi_6(\infty)$ is the steady-state value after the system has already phase-separated.(b) and (c) shows the temporal scaling using functions $t\to t\alpha^{2.8}$ and $t \to t e^{14r_c}$ respectively, which leads to a data collapse in both the cases.}
\label{figs7}
\end{figure}

In Fig.~3(d) and (e) of the maintext, we studied the dynamics of phase ordering by plotting ${\psi_6(t)}/{\psi_6(\infty)}$ as a function of time for different $\alpha$. We noticed that the system exhibit a delayed transition which scales as $t\to r_c^{8.5}t$. We note two important points here: (1) The temporal scaling is applied using $r_c$ since it is an intrinsic length-scale and can be obtained in general for any form of the potential used. We show in Fig.~\ref{figs7}(b), that the same collapse can be obtained by using a scaling of the form $t\to \alpha^{2.8}t$. (2) The large value of the scaling exponent $\lambda=8.5$, mainly indicates the sensitive dependence of the transition time on $\alpha$; the underlying dynamics may not be necessarily scale-free. The large exponent value may also imply that this dependence is exponential. To show this, we use an exponential scaling of the form $t\to e^{14 r_c} t$ (see Fig.~\ref{figs7}(c)) which also leads to a very good collapse of the data.
 
\section{MSD Plots at low density}
\label{AppendixG}
\begin{figure}[h!]
\centering
\includegraphics[width=0.25\textwidth]{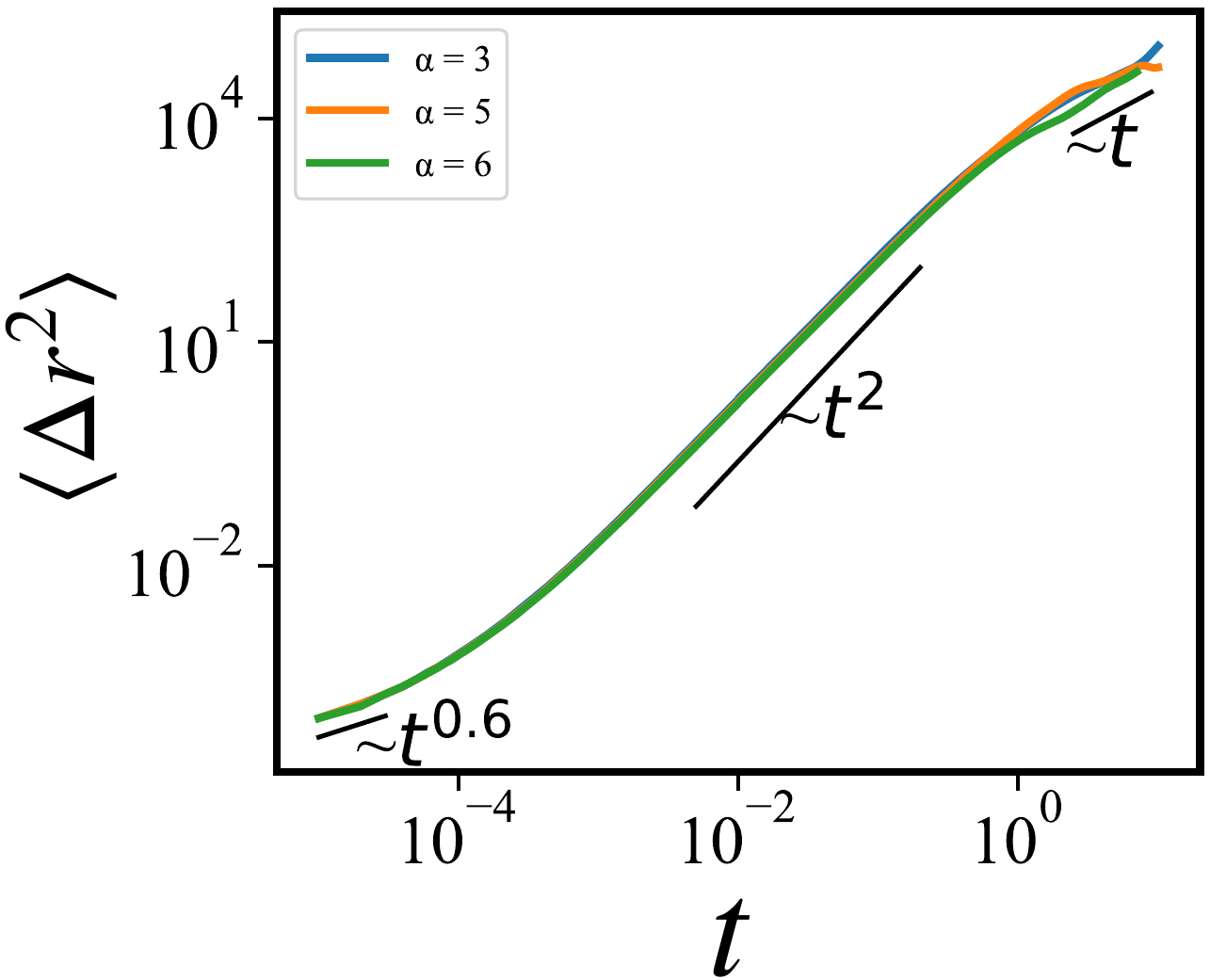}
\caption{Mean-square displacement for the non phase separating case at $\phi = 0.30 $ and Pe=150 for $\alpha=3,5$ and 6 respectively. In all the cases we notice a crossover from sub-diffusion$\to$ballistic$\to$diffusion.}
\label{figs8}
\end{figure}
Unlike at high density, as discussed in the maintext (Fig.~4), the system exhibit ballistic transport at intermediate time scales, when the system is kept a low density ($\phi\leq 0.30$), independent of the value of for different $\alpha$. This is shown in Fig.~\ref{figs8}. In the long-time limit, the dynamics again becomes diffusive as expected. At short time scales, the dynamics is driven by thermal fluctuations but sub-diffusive due to interparticle interactions.

\section{Simulation movie}
\label{AppendixH}
Movie~1: Steady state dynamics for overlap parameter $\alpha=3$ at Pe=150 and $\phi=0.7$. Although the system does not phase separate, the dynamics is not uniform and exhibit formation of local transient clusters.

\bibliography{scn_pre_rapid}

\begin{thebibliography}{42}
\bibitem [{foo()}]{footnote}%
  \BibitemOpen
  \href@noop {} {\bibinfo  {journal} {In Eq.~4, the main $\alpha$ dependence
  comes from the terms other than $C$, so it is reasonable to assume $C$ to be
  a constant}\ }\BibitemShut {NoStop}%
\end{thebibliography}
\end{document}